\documentclass[final,5p,times,twocolumn]{elsarticle}

\usepackage{lineno}
\usepackage[hidelinks]{hyperref}

\usepackage{graphicx}
\usepackage{amsmath}  
\usepackage{booktabs}
\usepackage{caption}
\usepackage{subcaption}
\usepackage{comment}

\def\pmbanner{{\hrule height 1 pt}\vskip35pt{NIMA POST-PROCESS BANNER TO BE REMOVED AFTER FINAL ACCEPTANCE}\vskip35pt{\hrule height 4pt}\vskip20pt}

\begin{document}

\begin{frontmatter}

\title{\pmbanner Lightweight Thermal Management Strategies for the Silicon Detectors of CBM at FAIR}

\author[1]{Franz Matejcek\corref{cor1}%
    \fnref{fn1}}
\ead{matejcek@physik.uni-frankfurt.de}

\author[2]{Kshitij Agarwal
    \fnref{fn1}}

\author[]{for the CBM Collaboration}

 \cortext[cor1]{Corresponding author}
 \fntext[fn1]{Joint first authors with equal contribution.}

\affiliation[1]{organization={Goethe-Universität Frankfurt}, 
                 addressline={Institut für Kernphysik, Max-von-Laue-Straße 1},
                 postcode={60438}, 
                 city={Frankfurt am Main}, 
                 country={Germany}}
\affiliation[2]{organization={Eberhard-Karls-Universität Tübingen}, 
                 addressline={Physikalisches Institut, Auf der Morgenstelle 14},
                 postcode={72076}, 
                 city={Tübingen}, 
                 country={Germany}}

\begin{abstract}
Lightweight thermal management is central to the design of the all-silicon Inner Tracker of the Compressed Baryonic Matter Experiment (CBM) at the Facility for Antiproton and Ion Research (FAIR). This experiment aims to study strongly interacting matter at neutron star core densities through fixed-target Au-Au collisions ($\sqrt{s_{NN}} = 2.9 - 4.9$\,GeV; up to 10$^7$ beam-target interactions per second; polar angular acceptance of $2.5^{\circ} \leq \Theta \leq 25^{\circ}$). Studies with thermal demonstrators for both the pixel-based Micro Vertex Detector and the strip-based Silicon Tracking System have validated their thermal management strategies, preparing them for series production now and global commissioning in 2028.
\end{abstract}


\end{frontmatter}

\section{Introduction}
\label{sec:intro}
The all-silicon Inner Tracker of CBM~\cite{CBM:2016kpk,Agarwal:2023otg}, comprising the Micro Vertex Detector (MVD)~\cite{Klaus:246516} and Silicon Tracking System (STS)~\cite{Heuser:2013nft, Heuser:2024adt}, is tasked to accurately determine secondary vertices and reconstruct tracks in a wide momentum range down to a few 100\,MeV/c in a high track-density environment resulting from unprecedentedly high beam-target interaction rates (see Tab.~\ref{tab:idprop}). This is expected to result in maximum accumulated non-ionising damage of up to $10^{14}\,n_{eq}/\textrm{cm}^2$ over the detector's lifetime. Lightweight thermal management strategies are crucial to mitigate radiation-induced effects, such as the rise of leakage currents over the detector's lifetime, while also minimizing the material budget to guarantee the required precision in momentum determination.  

\begin{table}[!t]
\small
\begin{tabular}{@{}lcc@{}}
\toprule
& \begin{tabular}[c]{@{}c@{}}MVD\end{tabular} 
& \begin{tabular}[c]{@{}c@{}}STS\end{tabular} \\ 
\midrule \midrule

\begin{tabular}[c]{@{}l@{}}Number of \\ Layers\end{tabular} 
& \begin{tabular}[c]{@{}c@{}}4 (8 – 20\,cm \\ Downstream) \\ In Vacuum\end{tabular} 
& \begin{tabular}[c]{@{}c@{}}8 (30 – 100\,cm \\ Downstream) \\ In Air\end{tabular} \\ \midrule

\begin{tabular}[c]{@{}l@{}}Active Area and \\ Granularity\end{tabular} 
& \begin{tabular}[c]{@{}c@{}}0.15 m$^2$ (288 Sensors)\\ 150M Pixels\end{tabular} 
& \begin{tabular}[c]{@{}c@{}}4 m$^2$ (876 Sensors)\\ 1.8M Strips\end{tabular} \\ \midrule

\begin{tabular}[c]{@{}l@{}}Sensor\\ Technology\end{tabular} 
& \begin{tabular}[c]{@{}c@{}}CMOS MAPS \\ (MIMOSIS)\end{tabular} 
& \begin{tabular}[c]{@{}c@{}}Double-Side Strips \\ (Hamamatsu Sensors)\end{tabular} \\ \midrule

\begin{tabular}[c]{@{}l@{}}Material per Layer\end{tabular} 
& \begin{tabular}[c]{@{}c@{}}0.3\% – 0.5\%\,X$_0$ \end{tabular} 
& \begin{tabular}[c]{@{}c@{}}0.3\% – 2\%\,X$_0$ \end{tabular} \\ \midrule

\begin{tabular}[c]{@{}l@{}}Spat. Res. $\sigma_{xy}$\\ Time Res. $\sigma_{t}$\end{tabular} 
& \begin{tabular}[c]{@{}c@{}}$\approx$ 5\,\textmu m \\ 5\,\textmu s (Frame Length)\end{tabular} 
& \begin{tabular}[c]{@{}c@{}}$\approx$ 10\,\textmu m \\ 5\,ns\end{tabular} \\ \midrule

\begin{tabular}[c]{@{}l@{}}Power\\ Dissipation\end{tabular} 
& \begin{tabular}[c]{@{}c@{}}$P_{\text{Sensor}}~\lesssim$~75\,mW/cm$^2$ \\$T_{\text{Sensor}}~\lesssim 0\,^{\circ}$C \\$\Delta T_{\text{Quadrant}}~< 10\,$K\end{tabular} 
& \begin{tabular}[c]{@{}c@{}}$P_{\text{Sensor}}~\approx$~50\,mW/cm$^2$ \\ $P_{\text{FEE}}~\approx$~40\,kW \\$T_{\text{Sensor}}~\approx 10\,^{\circ}$C\end{tabular} \\ \midrule

Rate & \begin{tabular}[c]{@{}c@{}}80\,MHz/cm²\\ 0.1\,MHz Reactions\end{tabular} & \begin{tabular}[c]{@{}c@{}}10\,MHz/cm² \\ 10\,MHz Reactions\end{tabular}   \\ \bottomrule
\end{tabular}
\caption{Summary of major features of MVD and STS.}
\label{tab:idprop}
\end{table}

\section{Thermal Management Concept}
\label{sec:concept}

\subsection{Micro Vertex Detector (MVD)}
\label{sec:concept_mvd}

\begin{figure}[!t]
     \centering
     \begin{subfigure}[b]{0.475\textwidth}
         \centering
         \includegraphics[angle = 270, width=8.6cm, trim = 0 0 0 0, clip]{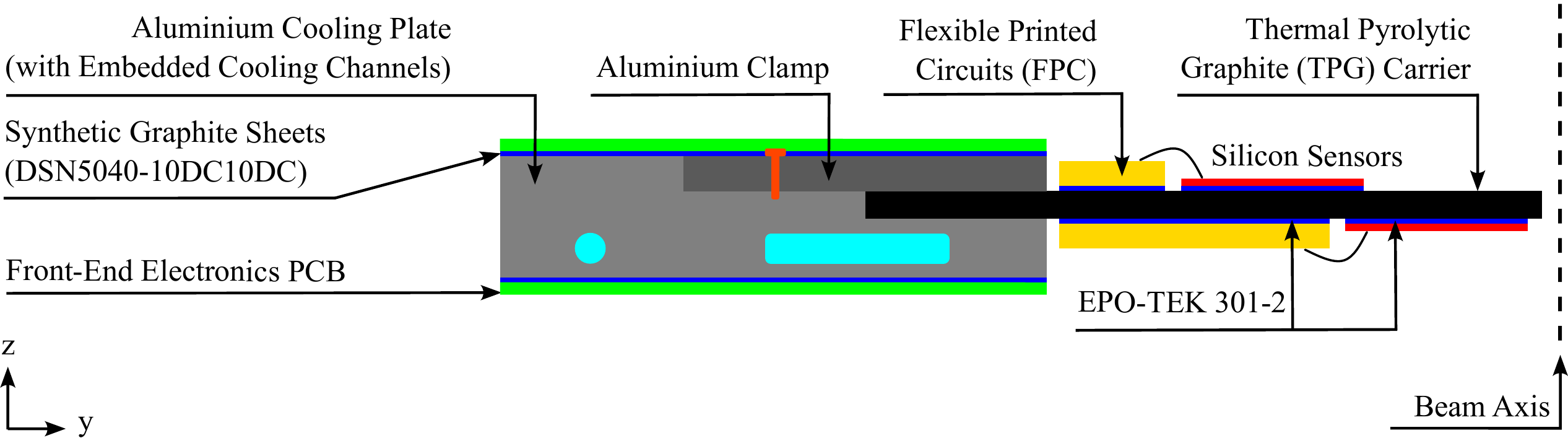}
         \caption{}
         \label{fig:mvd_illustration_a}
     \end{subfigure}
     \hfill
     \begin{subfigure}[b]{0.2375\textwidth}
         \centering
         \includegraphics[angle = 0, width=4.2cm, trim = 0 0 0 0, clip]{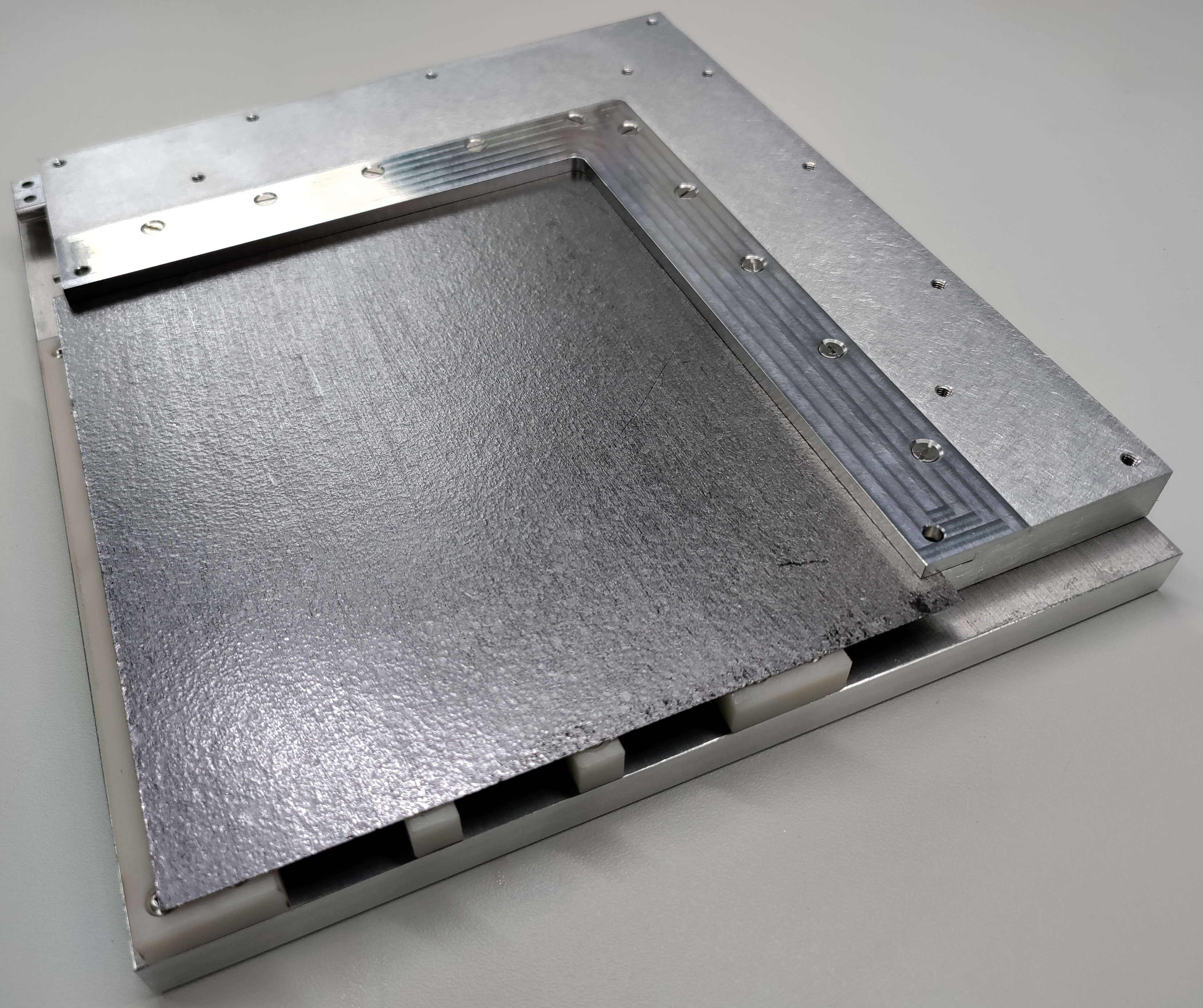}
         \caption{}
         \label{fig:mvd_illustration_b}
     \end{subfigure}
     \hfill
     \begin{subfigure}[b]{0.2375\textwidth}
         \centering
         \raisebox{0.88cm}[0cm]{\includegraphics[angle = 0, width=4.2cm, trim = 0 0 0 0, clip]{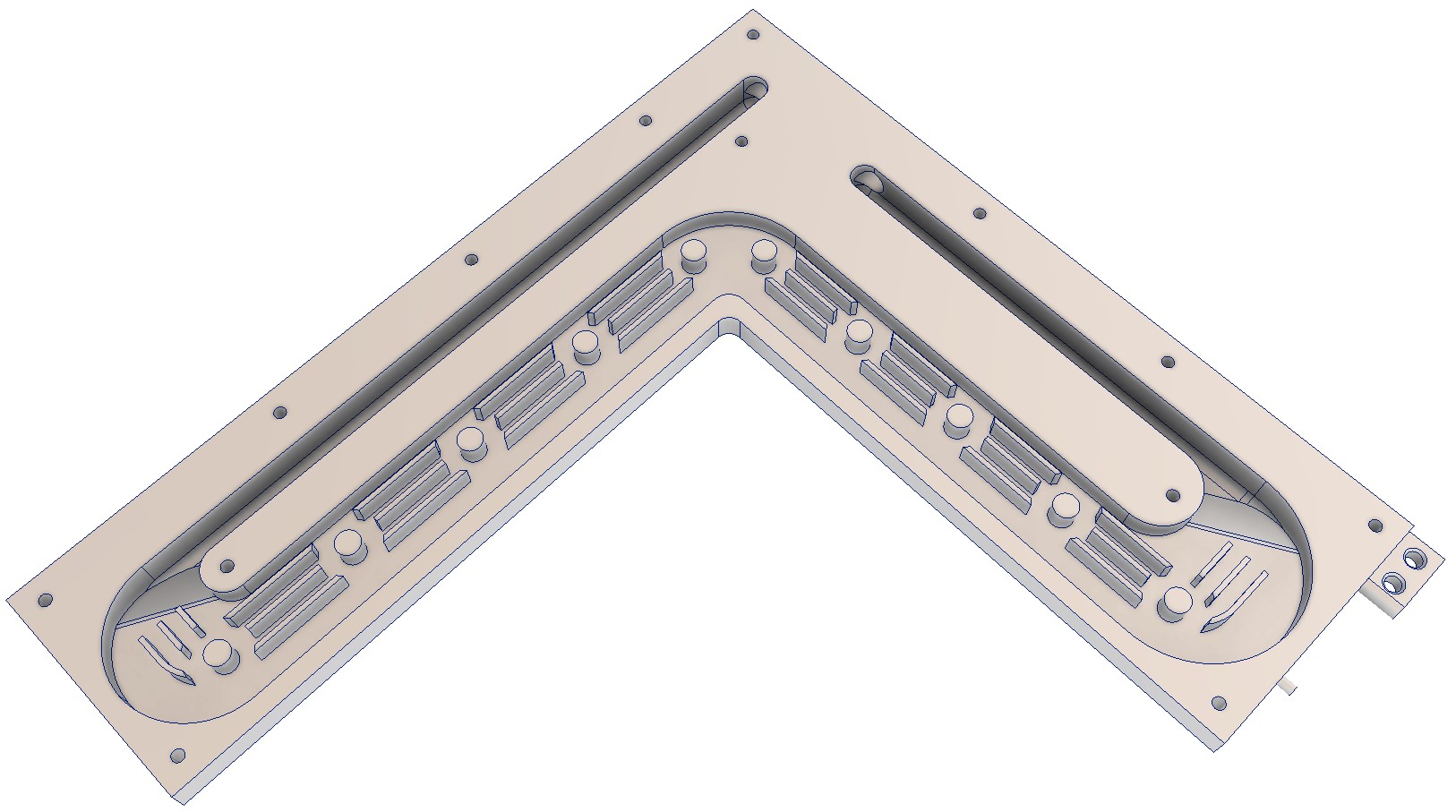}}
         \caption{}
         \label{fig:mvd_illustration_c}
     \end{subfigure}
        \caption{(a) Illustration showing the thermal path of the MVD (drawing is not to scale. (b) MVD quadrant with only the TPG carrier and aluminium heat sink. (c) CAD rendering of the milled fluidic channels inside the heat sink.}
        \label{fig:mvd_illustrations}
\end{figure}

Due to the vacuum operation of the MVD, the power dissipated by the sensors (MIMOSIS, $\lesssim$~75\,mW/cm$^2$, $\approx$\,120\,W total) is transported conductively through 380\,\textmu m thick Thermal Pyrolytic Graphite (TPG) sheets ($\lambda_{\text{in plane}}\gtrsim$\,1500\,W/m/K). The heat is guided into the peripherally located aluminium heat sinks which are actively cooled with liquid 3M\textsuperscript{TM} Novec\textsuperscript{TM}~649 (see Fig.~\ref{fig:mvd_illustrations}). This minimizes the material budget in the geometric acceptance and allows for a  homogeneous temperature distribution on the sensors ($<$10\,K over the TPG), ensuring a high detection efficiency and low fake hit~rate. The detector features a modular design with four stations assembled from four quadrants each. The material budget inside the acceptance (0.3\% – 0.5\%\,X$_0$ per station) is inhomogeneous due to overlapping sensors and FPCs on the double-side integrated TPG carriers. 

\subsection{Silicon Tracking System (STS)}
\label{sec:concept_sts}

The basic building block of the STS is called the module, comprising 320\,\textmu m thick double-sided micro-strip silicon sensors connected to a pair of front-end electronics (FEE) boards via up to 500\,mm long ultra-thin aluminium-polyimide cables. Modules are sequentially stacked over carbon-fibre support structures, called ladders, with the microcables dominating the material budget ($\lesssim 0.3\%$\,X$_0$ per microcable per sensor module). A total of 876 modules are distributed over 106 ladders across 8 tracking layers to optimize the material budget requirements, ensuring a momentum resolution $\delta p/p \leq 2\%$. 

\begin{figure}[!t]
     \begin{subfigure}[b]{0.475\textwidth}
         \centering
         \includegraphics[angle = 0, width=8.7cm, trim = 0 0 0 0, clip]{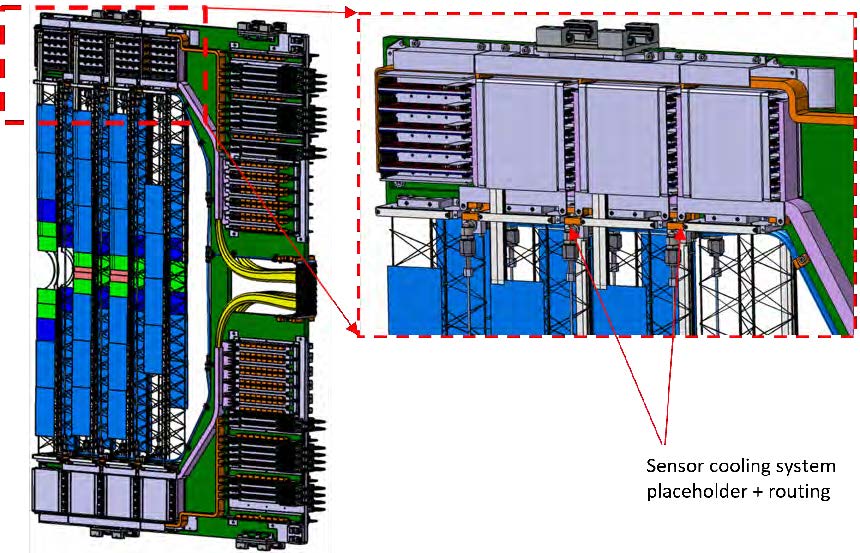}
         \caption{}
         \label{fig:sts_illustration_b}
     \end{subfigure}
     \hfill
     \begin{subfigure}[b]{0.475\textwidth}
         \centering
         \includegraphics[angle = 270, width=8.7cm, trim = 0 0 0 0, clip]{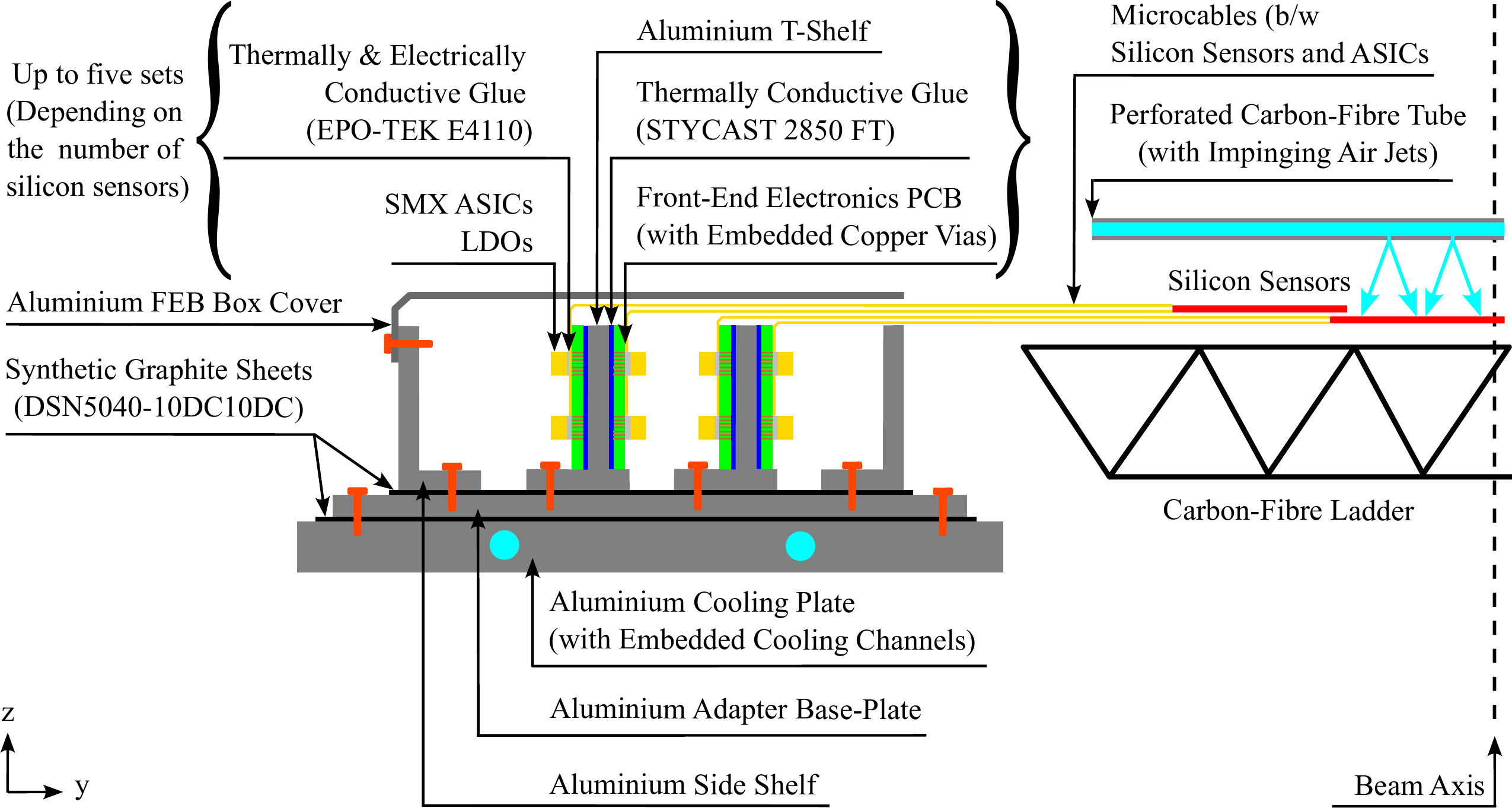}
         \caption{}
         \label{fig:sts_illustration_a}
     \end{subfigure}
     \hfill
     \begin{subfigure}[b]{0.475\textwidth}
         \centering
         \includegraphics[angle = 0, width=8.7cm, trim = 0 0 0 0, clip]{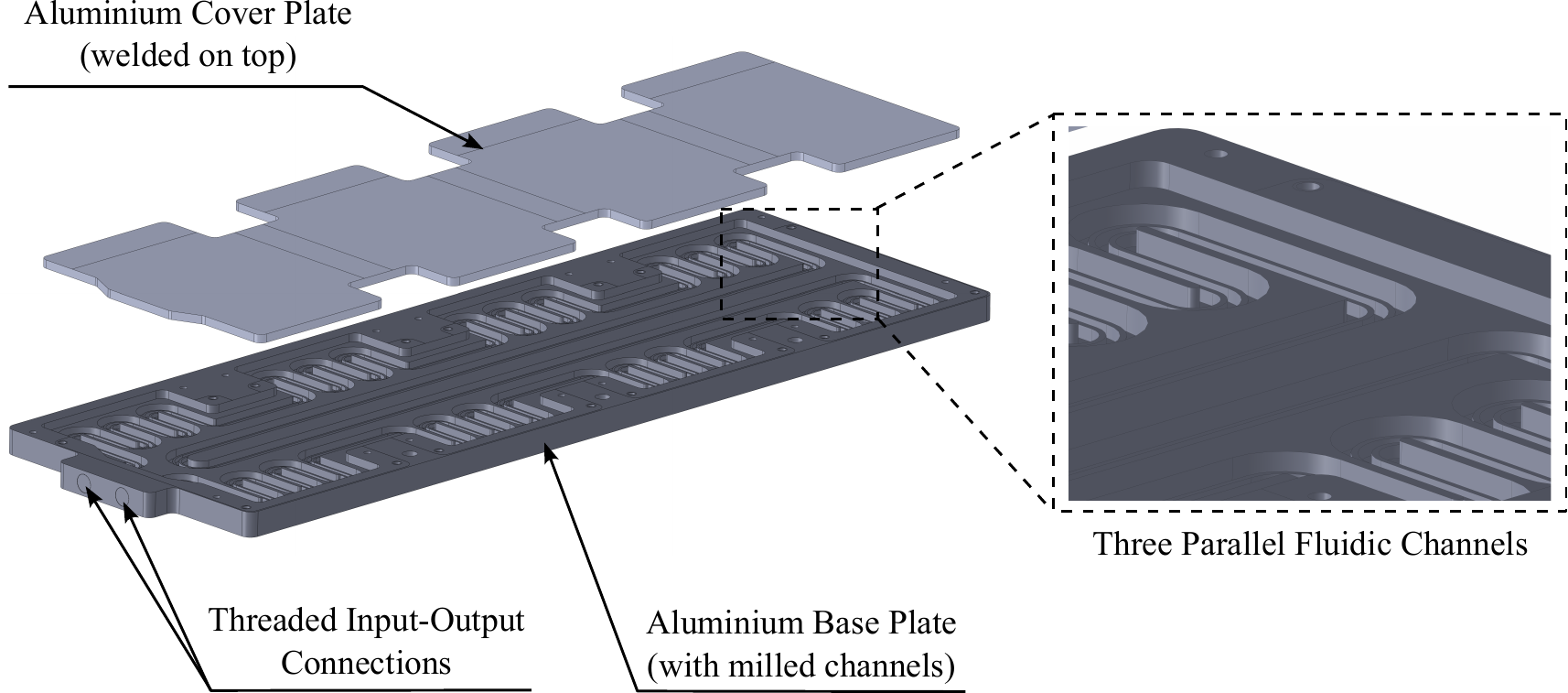}
         \caption{}
         \label{fig:sts_illustration_c}
     \end{subfigure}
        \caption{(a) CAD rendering of an assembled STS half-unit, with the call-out showing the perforated CF-tubes. (b) Illustration showing the thermal path of the STS-FEE (drawings is not to scale). (c) CAD rendering of the cooling plate with its comprising components and the zoomed-in view of the fluidic channels.}
        \label{fig:sts_illustrations}
\end{figure}

The silicon sensors around the beam-pipe are exposed to the most radiation and consequently dissipate the highest power ($\approx$~50\,mW/cm$^2$ at 10\,$^{\circ}$C and $10^{14}\,n_{eq}/\textrm{cm}^2$) due to the fixed-target detector geometry. These sensors are cooled by impinging air-jets via perforated carbon-fibre tubes, while natural air convection is deemed sufficient for the less irradiated sensors (see Fig.~\ref{fig:sts_illustration_b}). This avoids their thermal runaway by maintaining a stable operating temperature of $\approx 10\,^{\circ}$C for optimal signal-to-noise ratio ($S/N \geq 10$) and avoids reverse annealing
(full depletion voltage $V_{dep} < 500$\,V), while keeping a minimum material budget inside the acceptance. 

The FEE power dissipation ($\approx$\,40\,kW) is neutralized through a heat-conductive path with minimal thermal impedance into the friction-stir welded cooling plate which is cooled with mono-phase liquid 3M\textsuperscript{TM} Novec\textsuperscript{TM}~649 (see Figs.~\ref{fig:sts_illustration_a}-\ref{fig:sts_illustration_c}) to minimize the residual heat-transfer to the silicon sensors. 

\section{Experimental Validation}
\label{sec:exp}

Realistic demonstrators with pre-series detector components were built for both subsystems to validate the respective thermal management strategies, explore safety margins, and determine optimal operational parameters. 

\subsection{Micro Vertex Detector (MVD)}
\label{sec:exp_mvd}

\begin{figure}[!t]
     \centering
     \includegraphics[width=0.35\textwidth]{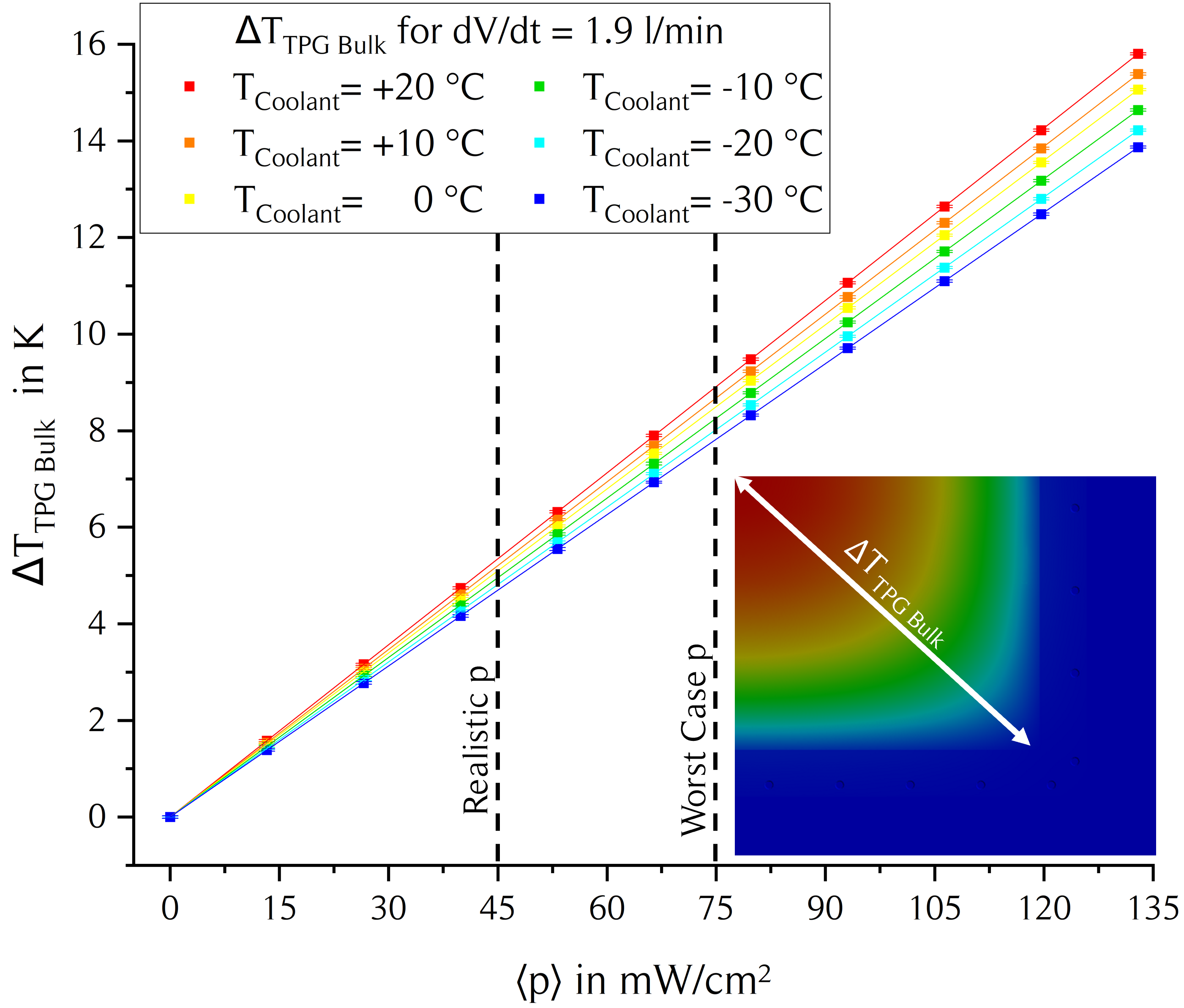}
     \caption{Measured maximum temperature difference across the TPG carrier as a function of the sensor's power dissipation and the coolant temperature.}
     \label{fig:mvd_results}
\end{figure}

The MVD demonstrator module, comprising an L-shaped heat sink and a TPG carrier in a vacuum chamber, represents the entire detector with its 16 thermally independent quadrants. Fig.~\ref{fig:mvd_results} shows the radiation-corrected maximum temperature difference across the TPG bulk of an MVD quadrant ($\Delta T_{\textrm{TPG~Bulk}}$) for varying sensor power and coolant temperatures. The radiative heat entry from the surroundings (0\,mW/cm$^2$ for $T_{\text{Coolant}}=+20\,^{\circ}$C, up to 20\,mW/cm$^2$ for $T_{\text{Coolant}}=-30\,^{\circ}$C) has been measured and subsequently subtracted from the data. The data shows that $\Delta T_{\textrm{TPG~Bulk}} <$~10\,K at nominal operational conditions with large safety margins over a wide range of coolant temperatures. The thin thermal interface layer of $\sim$\,30\,\textmu m between TPG and sensors ensures a negligible temperature differences between the two. The requirements of temperature difference and operation temperature (see Tab.~\ref{tab:idprop}) can be met. 

\subsection{Silicon Tracking System (STS)}
\label{sec:exp_sts}

\begin{figure}[!t]
     \centering
     \begin{subfigure}[b]{0.38\textwidth}
         \centering
         \includegraphics[width=\textwidth]{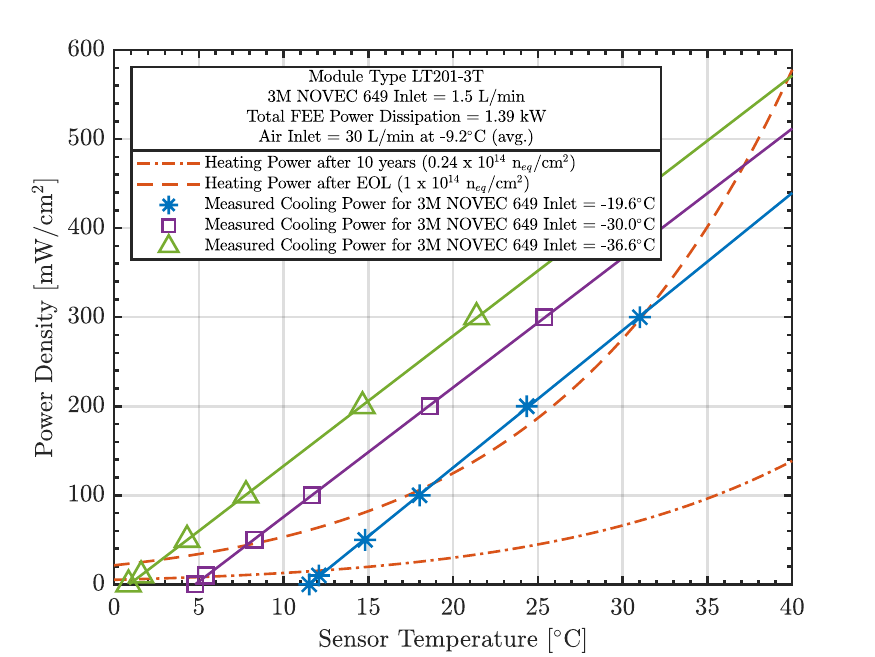}
         \caption{}
         \label{fig:sts_results_a}
     \end{subfigure}
     \hfill
     \begin{subfigure}[b]{0.45\textwidth}
         \centering
         \includegraphics[width=\textwidth]{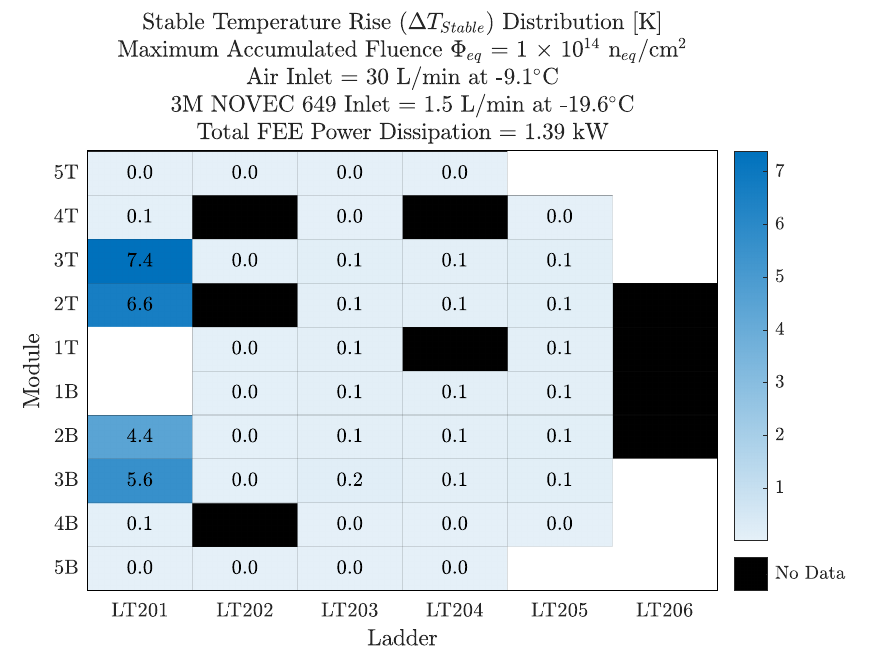}
         \caption{}
         \label{fig:sts_results_b}
     \end{subfigure}
        \caption{(a) Thermal runaway behavior for the most irradiated module cooled by impinging air jets for various 3M\textsuperscript{TM}~Novec\textsuperscript{TM}~649 inlet temperatures, after 10 years and EOL operation. (b) Temperature rise of all half-station's sensors before and after power dissipation mimicking the EOL fluence distribution for baseline operational parameters.}
        \label{fig:sts_results}
\end{figure}

The STS thermal demonstrator comprises one thermally active half-layer with dummy silicon sensor and electronics heating elements in realistic mechanical boundary conditions to accurately represent the interplay between the heating elements. The experimental setup is cooled by pilot cooling plants for the FEE with 3M\textsuperscript{TM}~Novec\textsuperscript{TM}~649, and silicon sensors with cold and dry air. The baseline cooling parameters are determined by numerical simulations of the cooling concept and the survival properties of the STS components under extended cooling conditions. This includes FEE and thermal interface thermal cycling, and sensor vibrations under air flow, providing limits of $-20\,^{\circ}$C 3M\textsuperscript{TM}~Novec\textsuperscript{TM}~649 temperature and 30\,L/min air flow rate, respectively. Stable sensor temperatures ($T_{\textrm{Stable}}$) of 12.6\,$^{\circ}$C and 18.9\,$^{\circ}$C are observed for the hottest sensor dummy that correspond to a power dissipation after accumulating fluences for 10 and 40 years (end-of-lifetime, EOL) operation, respectively. Sufficient margins from the thermal runaway are respected (see Fig.~\ref{fig:sts_results_a}). The rise of the stable temperature ($\Delta T_{\textrm{Stable}}$) observed with sensor power dissipations corresponding to the EOL fluence distribution exhibits the importance of the air cooling by both forced air convection via impinging air jets (central ladders; LT201-202; $\Delta T_{\textrm{Stable}} <$~7.4\,K) and natural convection (peripheral ladders; LT203-206; $\Delta T_{\textrm{Stable}} \approx$\,0\,K); all while neutralising the FEE power dissipation (see Fig.~\ref{fig:sts_results_b}). Collectively, it can seen that $T_{Sensor} \approx 10\,^{\circ}$C can be maintained after 10 years of operation with the baseline 3M\textsuperscript{TM}~Novec\textsuperscript{TM}~649 inlet temperature of $-20\,^{\circ}$C. For longer detector operation up to EOL fluence, the 3M\textsuperscript{TM}~Novec\textsuperscript{TM}~649 inlet temperature can be lowered down to obtain $T_{Sensor} \leq 10^{\circ}$C. 

\section{Acknowledgements and Funding}
\label{sec:funding}

FM and KA acknowledge the support from the BMBF Project-ID 05P21RFFC2 and 05P19VTFC1, respectively.

\bibliographystyle{elsarticle-num-names} 
\bibliography{bibfileTemplate}

\end{document}